\documentstyle[preprint,version2,aps]{revtex}
\draft
\begin{document}

\begin{title}
Spin Dynamics in the Normal State of  High ${\rm T_c}$ Superconductors
\end{title}

\author{Qimiao Si}
\begin{instit}
Serin Physics Laboratory, Rutgers University, Piscataway, NJ 08855-0849, USA
\end{instit}

\begin{abstract}

We summarize our recent theoretical studies on the spin dynamics in
the normal state of the metallic cuprates. The contrasting wave vector
dependence of the dynamical spin structure factor $S({\bf q}, \omega
)$ in LaSrCuO and YBaCuO systems are attributed to the differences in
the fermiology, in conjunction with strong Coulomb correlations. These
effects are found to account also for the anomalous temperature and
frequency dependence of $S({\bf q}, \omega )$. We conclude that the
low energy spin dynamics of the metallic cuprates are described in
terms of correlated quasiparticles with a Luttinger Fermi surface and
a non-zero antiferromagnetic exchange interaction.

\end{abstract}

{\it Invited Review, to appear in Int. Journ. Mod. Phys. B}

\newpage

\section{\bf Introduction}
\label{sec:intro}

In the past few years, the study of various physical properties have
provided many insights into the nature of both the normal and the
superconducting states of high ${\rm T_c}$ copper
oxides.\cite{Ginsberg,Bedell} We discuss here the spin dynamics in the
metallic cuprates,\cite{Birge,Slichter} with a focus on the normal
state. The systematic study of the spin dynamics is motivated in part
by the widespread belief that they may be relevant to the
superconductivity. It is also of importance in elucidating the nature
of doped Mott insulators in general, a  fundamental and long standing
problem in condensed matter physics.

The spin dynamics in the cuprates were studied first in the undoped
compounds, such as ${\rm La_2CuO_4}$ and ${\rm YBa_2Cu_3O_6}$. The
electronic structure in these systems is characterized by the
half-filling of the outermost (copper) band. According to the one
electron picture, these half-filled cuprates should be metals.
However, they all show three-dimensional magnetic ordering at low
temperatures,\cite{Birge} and  continue to be insulators above the
N\'eel temperature.\cite{Cheong} In this paramagnetic insulating
phase, spin dynamics have been studied at low energies through
inelastic neutron scattering, and at higher energies from both
(two-magnon) Raman scattering\cite{Raman} and (zone-boundary
one-magnon) neutron-scattering\cite{Hayden}. These results have led to
a fairly convincing picture that,\cite{Chakra} the paramagnetic phase
is described by the two-dimensional spin ${ 1\over 2}$ Heisenberg
model defined on the copper square lattice within a ${\rm CuO_2}$
layer. Combined with the existence of a large charge gap,\cite{Cheong}
this indicates that the half-filled cuprates are Mott
(Charge-transfer) insulators: the electrons are localized due to the
strong on-site Coulomb repulsions, and exhibit at low energies spin
degrees of freedom only.

As the system is doped away from half filling, the magnetic ordering
quickly vanishes, and an insulator to metal (superconductor)
transition sets in. Our understanding of the half-filled case does
not, however, uniquely specify the nature of the doped metallic phase.
Many questions can be asked about the role of doping. For example, 1)
doped holes disturb the spin background. Do they lead to a frustrated
spin system, or manage to change the nature of the spins altogether?
The observation of a Luttinger Fermi surface in the metallic cuprates
implies that, the localized copper spins of the half-filled limit are
converted to itinerant electrons at low energies in the doped regime.
How is this manifested in the spin dynamics? 2) the Mott insulating
nature of the half-filled case suggests that, the cuprates have strong
Coulomb correlations. Is this relevant to the understanding of the low
energy spin dynamics in the doped metallic phase? 3) in the
half-filled case, the magnetic interactions are mainly between the
nearest neighbor copper spins. What is the nature and the magnitude of
the effective exchange coupling between low energy spin excitations in
the doped metallic phase? Determining this coupling is of course
important for the purposes of addressing the relevance of spin
fluctuations to superconductivity; and 4) in the Mott insulating
state, charge excitations are pushed to high energies while spin
excitations remain at low energies. How do the doped holes modify the
coupling between charge and spin degrees of freedom?

In the metallic cuprates, a host of anomalies in the low energy spin
dynamics have been revealed in nuclear magnetic/quadrupolar resonance
(NMR/NQR) and inelastic neutron scattering experiments. One striking
feature is the strong contrast between the spin fluctuation spectra,
$S({\bf q}, \omega )$, in LaSrCuO and YBaCuO families. Other anomalies
are observed in the wave vector, temperature and frequency dependence
of $S({\bf q}, \omega )$. In the following, we will demonstrate that
the low energy particle-hole spin excitations near the Luttinger Fermi
surface can naturally explain the differences in $S({\bf q}, \omega )$
between LaSrCuO and YBaCuO systems. At the same time, strong Coulomb
correlations play an important role for understanding these magnetic
data. Our analysis also establishes the existence of a non-zero
antiferromagnetic exchange coupling between the quasiparticles.
Finally, we will briefly comment on the relation between low energy
spin and charge dynamics.

The rest of the paper is organized as follows. In section
\ref{sec:data} we give a brief overview of the spin dynamical data in
the metallic cuprates. Section \ref{sec:forma} presents a formalism
for magnetic interactions and dynamical spin susceptibility within the
extended Hubbard model in the limit of strong Coulomb correlations.
This formalism is used to analyze the spin fluctuation spectra in
Section \ref{sec:theory}. Section \ref{sec:discu} contains some brief
discussion on the spin dynamics in the superconducting state and the
relation between spin and charge dynamics. Several concluding remarks
are made in Section \ref{sec:conclu}.

This brief review summarizes mainly our own
work,\cite{Magint,Lu,Zha1,Si} and is not intended to be comprehensive.
Readers are directed to other reviews for more discussions on the
subject.\cite{Reviews} A more complete analysis of the normal state
properties in the metallic cuprates along similar lines can be found
in Ref. \cite{Review}, which also contains more extensive references.
The importance of fermiology in understanding the spin fluctuation
spectra in LaSrCuO has been emphasized independently by Littlewood
{\it et al.}\cite{Littlewood} in which correlation effects are treated
through inelastic quasiparticle lifetimes within the Marginal Fermi
liquid scheme. The role of fermiology in understanding the contrasting
spin dynamics in LaSrCuO and YBaCuO has also been emphasized
subsequently by Tanamoto {\it et al.}\cite{Fukuyama} within the gauge
theory approach to the t$-$J model. These studies have all led to the
conclusion that the bulk of the low energy spin fluctuation spectra
are described in terms of itinerant renormalized quasiparticles, with
no additional exotic dynamics in the spin channel.

\bigskip
\section{\bf Magnetic Data in the Metallic Cuprates}
\label{sec:data}

Both NMR/NQR spin-lattice relaxation rate and the inelastic neutron
scattering cross section measure the dynamical spin structure factor
$S({\bf q}, \omega)$. From the fluctuation-dissipation theorem,

\begin{eqnarray}
S({\bf q}, \omega) ~=~ ( 1 + n_b (\omega ) ) {\rm Im} \chi ({\bf q},
\omega)
\label{flu-dis}
\end{eqnarray}
where $\chi ({\bf q}, \omega)$ is the transverse dynamical  spin
susceptibility, and $n_b(\omega)=1/(e^{\omega/T}-1)$ is a Bose factor.

The spin-lattice relaxation rate, ${1\over T_1}$, is the rate at which
the nuclear magnetization relaxes towards equilibrium starting from a
non-equilibrium distribution. Such a spin flipping relaxation is
induced by the electronic spin excitations, through a hyperfine
coupling between the nucleus and the electrons. Specifically,

\begin{eqnarray}
({1\over T_1})_r ~\sim ~ \sum_{\bf q} A_r^2({\bf q}) S({\bf q},
\omega_o\rightarrow 0)
\label{t1}
\end{eqnarray}
\noindent
Since the nuclear resonance frequency $\omega_o$ is essentially zero
in the scale of electronic energies, ${1\over T_1}$ measures spin
dynamics at the vanishing frequencies. Due to the local nature of the
nucleus dynamics, ${1 \over T_1}$ measures the wave vector ${\bf
q}$-integrated $S({\bf q},\omega_o)$, weighted by the ${\bf
q}$-dependent hyperfine coupling constant $A({\bf q})$. Qualitative
information about the ${\bf q}$-dependence in $S({\bf q},\omega_o)$
can be derived if there exist several kinds of nuclei at different
sites (labeled by $r$ in Eq. (\ref{t1})) which relax via the same
dynamical structure factor, providing the associated hyperfine
coupling $A_r({\bf q})$ has dominant contributions from different
regions in the ${\bf q}$-space for different sites $r$.

Inelastic neutron scattering experiments measure the dynamical spin
structure factor more directly. Here the cross section

\begin{eqnarray}
{d^2\sigma \over {d\Omega dE}} ~\sim~ S({\bf q}, \omega)
\label{neutron}
\end{eqnarray}
The fact that both the wave vector and the frequency are resolved is
of course an important advantage of these measurements as compared
with NMR/NQR experiments.

That the spin dynamics in the metallic cuprates are anomalous compared
to conventional metals has been discussed extensively from the
temperature dependence of ${1 \over T_1}$. To appreciate this
anomalous behavior, we note that ${1 \over T_1}$ within a conventional
metal can be loosely estimated as follows. The number of electrons
available to flip a nuclear spin is proportional to $T N(E_F)$, where
$N(E_F)$ is the density of states at the Fermi level. Each electron
has a contribution to the relaxation rate proportional to $N(E_F)$,
which measures the number of available final states. Therefore, ${1
\over T_1}$ has a linear temperature dependence. Furthermore, since
the static uniform spin susceptibility, $\chi$, is proportional to the
density of states $N(E_F)$, there exists the Korringa relation,

\begin{eqnarray}
{1 \over T_1 T \chi^2} = 1
\label{korringa}
\end{eqnarray}
(where dimensionless units are used.)

Shown in Fig. \ref{1t1expt} are the relaxation rates at both the
planar copper and oxygen sites in (nearly optimal doped) ${\rm
YBa_2CuO_{7-\delta}}$ and that for the planar copper site in ${\rm
La_2Sr_{0.15}CuO_4}$.\cite{Imai,Hammel,Slichter} Two
peculiar features are associated with $({1 \over T_1})_{Cu}$: a) it is
strongly enhanced compared to that in conventional metals; The nominal
Korringa ratio is at least one order of magnitude larger;\cite{Hammel}
and b) its temperature dependence deviates considerably from linear
behavior. Finally, comparing $({1 \over T_1})_{O}$ and $({1 \over
T_1})_{Cu}$ gives yet another important feature: c) $({1 \over
T_1})_{O}$ is almost linear in temperature, and has a nominal
Korringa ratio of order 1.

Within the ``standard'' one-spin-component interpretation, these
anomalies are understood as follows.\cite{Hammel,Shastry,Bulut,MMP,Lu}
In a ${\rm CuO_2}$ plane of the metallic cuprates, the spin degrees of
freedom are mainly associated with the planar copper sites. Strong
antiferromagnetic spin correlations are responsible for the excess
relaxation of the copper nuclear magnetization, leading to an enhanced
$({1 \over T_1})_{Cu}$. The amount of enhancement depends on
temperature. Therefore, $({1 \over T_1})_{Cu}$ deviates from the
Korringa linear in $T$ behavior. On the other hand, since the oxygen
sites are located at the mid-point between two copper sites, the
contributions to $({1 \over T_1})_O$ from neighboring copper spins
cancel with each other. The anomalies are therefore absent for $({1
\over T_1})_O$.

In momentum space, one can infer that this interpretation is based on
the assumption that the dynamical spin susceptibility is strongly
peaked at the antiferromagnetic wave vector, ${\bf Q}_{\rm AF}= ({\pi
\over a}, {\pi \over a})$. The existence or not of such sharp ${\bf
q}$-structure in the dynamical spin susceptibility can, of course, be
more directly established from inelastic neutron scattering
experiments at low energies. In the past, neutron scattering studies
have elucidated the nature of spin fluctuations in other strongly
correlated fermion systems such as liquid ${\rm {}^3 He}$ and heavy
fermion materials.\cite{HeHF}

The momentum dependence in the inelastic neutron scattering cross
sections in metallic ${\rm YBaCuO}$ and in ${\rm LaSrCuO}$ are shown
in Fig.
\ref{neutronexpt}.\cite{Mason,Thurston,Tranquada,Rossat-Mignod} The
main features can be summarized as follows: a) in YBaCuO,
antiferromagnetic spin fluctuations are essentially commensurate.
However, the peaks are rather broad, and the peak widths are nearly
temperature independent; b) in ${\rm LaSrCuO}$, spin fluctuations are
sharply peaked at incommensurate wave vectors ${\bf Q^*}$. The
incommensurability, defined as the deviation of ${\bf Q^*}$ from ${\bf
Q}_{\rm AF}$, quickly increases as the doping concentration is
increased as shown in Fig. \ref{neutronexpt}(c). This
incommensurability is much larger (by more than a factor of two) than
that expected within an itinerant description in terms of the one band
Hubbard model with nearest neighbor hopping only.

In addition to the ${\bf q}$-structure, the spin dynamics have also
been extensively probed as a function of temperature and frequency.
Within the lightly-doped non-superconducting ${\rm LaSrCuO}$
system\cite{Keimer} $\chi''({\bf q},\omega )$ is found to scale with
${\omega \over T}$, which is consistent with the Marginal Fermi liquid
ansatz.\cite{MFL} Systematic studies of the temperature and
frequency dependence of $\chi''({\bf q},\omega )$ at higher doping
concentrations, while still under way for the normal state of
superconducting LaSrCuO\cite{comm}, have however revealed low energy
scales in YBaCuO as is illustrated in Fig. \ref{neutrontw} and
discussed extensively in Refs. \cite{Tranquada,Rossat-Mignod}.

It is indeed intriguing that, the two families of cuprates show
contrasting ${\bf q}$-dependence in the dynamical spin susceptibility.
In the following, we show how this difference helps understand the
nature of the low lying spin excitations. We will also demonstrate
that the dynamical spin susceptibility revealed in inelastic neutron
scattering in both YBaCuO and LaSrCuO can not be easily reconciled
with the form-factor cancellation argument for $({1 \over T_1})_O$.

\bigskip
\section{\bf Strongly Correlated Quasiparticle Description\\
of the Magnetic Dynamics}
\label{sec:forma}

To understand the spin dynamical spectra, we need a microscopic
formalism for the dynamical spin susceptibility $\chi ({\bf q},
\omega)$. In the following, we present such a scheme within the
extended Hubbard model in the presence of strong Coulomb
correlations.\cite{Magint}

\subsection{The Extended Hubbard Model in the Large $U$ Limit}

We consider the extended Hubbard model defined in a
${\rm CuO_2}$ layer,

\begin{eqnarray}
{\rm H} = &&\sum_{i~ \sigma} \epsilon_d^o~
d_{i\sigma}^{\dagger} d_{i\sigma}
 ~+~\sum_{l~\sigma} ~\epsilon_p~ p_{l \sigma}^{\dagger} p_{l \sigma}^{}
 ~ + ~\sum_{ l_1 l_2 \sigma }~ t_{l_1 l_2}
(p_{l_1 \sigma}^{\dagger} p_{l_2 \sigma} + h.c. )\nonumber\\
&&+~ \sum_{ i~l~ \sigma } V_{pd} ( d_{i \sigma}^{\dagger}
p_{l \sigma}^{} ~+~ h.c. ) ~+~ \sum_{i} ~U~ d_{i \uparrow}^{\dagger}
d_{i \uparrow} d_{i \downarrow}^{\dagger} d_{i \downarrow}
\label{model}
\end{eqnarray}
Here $p$ and $d$ denote the oxygen ${\rm p_x}$, ${\rm p_y}$ and the
copper $d_{x^2 - y^2 }$ orbitals respectively. We include a
hybridization $ V_{pd}$ between nearest neighbor copper and oxygen
orbitals, a finite oxygen dispersion derived from oxygen-oxygen
hopping matrix element $t_{pp}$ (nearest neighbor) and $t_{pp}'$ (next
nearest neighbor).
The oxygen and ``bare'' copper levels are called $\epsilon_p$ and
$\epsilon_d^o$ respectively. The on-site Coulomb repulsion for
the copper is $U$. The parameters involved in the Hamiltonian
(\ref{model}) appropriate to the copper oxides have been extensively
studied.\cite{parameter} Because of the large number of
parameters involved, we presume that the derived parameters
only specify an appropriate range. Specific parameters within this
range will be chosen from phenomenological constraints to be discussed
shortly. Before doing this, we first give a qualitative discussion for
the case that the Coulomb repulsion $U$ and the level separation
$\epsilon_p-\epsilon_d^o$ are large.

For the half-filled case, the system is a Mott (Charge-transfer)
insulator. Because of the exclusion of the double occupancy, the magnetic
interaction has no static on-site (${\bf q}$-independent)
component. The dominant contribution comes from the nearest neighbor
term. The low energy effective Hamiltonian is the spin
${1\over 2}$ Heisenberg model. Formally,

\begin{eqnarray}
J( {\bf q}) = J_o (cos(q_x) + cos(q_y))
\label{jq}
\end{eqnarray}
where the nearest neighbor superexchange interaction $J_o$ can be
derived from the full model (\ref{model}) through integrating out high
energy dynamics. The relevant high energy states is mainly oxygen in
character. This nearest-neighbor form is established to a good
accuracy through high energy spin wave measurement.\cite{Hayden}

In principle, there are many possible phases to which a Mott insulator
may evolve when doped. The study of these different fixed points and
the associated quantum phase transitions is a fundamental and unsolved
problem, which is beyond the scope of the present paper. We consider
here the scenario that the electron localization and the
antiferromagnetic ordering in the half-filled Mott insulator evolve
into two corresponding incipient instabilities when
doped.\cite{Review} The incipient electron localization is
characterized by an enhanced quasiparticle mass, while the incipient
magnetic instability is manifested as an effective exchange coupling
between the quasiparticles. Phenomenologically, such a picture is
consistent with the existence of a Luttinger Fermi surface, the small
and doping dependent plasma frequencies, and the presence of doping
dependent magnetic  fluctuations which we will analyze in detail.
Further support for this picture comes from a comparison with
anomalous physical properties in heavy fermions.\cite{Review,loc} This
picture is also consistent with indications from the small cluster
diagonalization studies that, doping leads to the formation of
quasiparticles with enhanced mass and a Luttinger Fermi
surface.\cite{Cluster} We emphasize that, our results are more general
than the particular approach we use to describe the model Hamiltonian
Eq. (\ref{model}). For our purposes here, this scheme can be viewed as
providing a handle to differentiate itinerant, and strongly Coulomb
renormalized, quasiparticle description of the  spin dynamics in the
metallic cuprates, versus other approaches.

Formally, both the incipient localization\cite{Kim,KLR,Newns} and the
magnetic interaction effects\cite{Magint} can be systematically
studied within the large N approach to the model Hamiltonian
(\ref{model}) (where N is the spin degeneracy). This approach has been
extensively used in the description of the Kondo lattice.\cite{Kondo}
For our purposes, the large N approach provides a convenient tool to
systematically study the doping dependence of magnetic properties. In
the end, it would be useful to compare our conclusions with numerical
analysis in a small cluster on generalized Hubbard models with
extended dispersion and strong Coulomb interactions.\cite{Moreo}

We choose the microscopic parameters of the model Hamiltonian
(\ref{model}) such that a small number of experimentally measurable
properties are fitted. These are the plasma frequencies and the Fermi
surface shapes. We then proceed to calculate the spin fluctuation
spectra and compare with experimental results.

\subsection{Renormalized Band Structure}

The band structure of the Coulomb renormalized quasiparticles is
described by the large N mean field theory. The mass enhancement is
determined microscopically by the renormalization of the hybridization
matrix element, $V_{pd} ~\rightarrow~V_{pd}^*$.
The renormalization of the hybridization in mean field theory is
mainly determined by the parameters associated with the copper states:
the bare hybridization $V_{pd}$ and the bare level difference
$\epsilon_p-\epsilon_d^0$ (U is taken as infinite). We choose these
bare parameters such that the renormalized plasma frequency, shown in
Fig. \ref{plasma}(a), fit the experimental values given in Fig.
\ref{plasma}(b) (derived from Drude fitting the optical conductivity
spectra\cite{optical}).

The second important feature of the renormalized band structure is
the shape of the Fermi surface. From the angular resolved
photoemission spectroscopy (ARPES) measurement\cite{ARPES} and the LDA
band structure calculations\cite{LDA}, the Fermi surface in
YBaCuO is found to be rotated by $45^o$ relative to that of a diamond
shape expected in a nearest neighbor tight binding form. The Fermi
surface in LaSrCuO retains a diamond-like shape, though it is somewhat
``twisted'' so that the nearly flat regions of the Fermi surface are
closer to the ${\rm \Gamma}$ point.\cite{LDA} The precise mechanism
for the difference in the Fermi surfaces in these systems is at
present not known. Since the spin degrees of freedom in the large $U$
limit are mainly associated with the copper states, within our model
study we choose appropriate oxygen dispersion to derive the
corresponding shape of the Fermi surface for each family.
Specifically, the Fermi surface ``twisting'' in LaSrCuO is derived by
choosing a nonzero $t_{pp}$, and the Fermi surface rotation in YBaCuO
by choosing a non-zero $t_{pp}'$ in addition to $t_{pp}$. These Fermi
surfaces are shown in Fig. \ref{fermisurface}.

The third feature associated with the renormalized band structure,
specifically related to the two dimensionality, is the
logarithmically divergent van Hove singularity in the density of
states.  The separation between the energy at which the van Hove
singularity occurs and the Fermi level, $\omega_{VH}=|E_{VH}-E_F|$,
will be shown to be manifested in the spin fluctuation spectra. In
this regard, we emphasize two important features. First, strong
Coulomb interactions pin the van Hove singularity near the Fermi level
over a wide range of doping concentrations.\cite{Resis,Newns2}
Secondly, due to the different shapes of the Fermi surfaces in YBaCuO
and LaSrCuO, the energy scale $\omega_{VH}$ for YBaCuO is larger
than that of LaSrCuO. Typically, we found  $\omega_{VH}$ of the
order $25 meV$ in YBaCuO, and less than $5meV$ in LaSrCuO.

\subsection{Magnetic Interactions and Dynamical Spin Susceptibility}

In the strong Coulomb correlation limit,
the quasiparticles are dominantly associated with the
copper states. Therefore, the exchange interaction between the
quasiparticles can be thought of as primarily between the copper
states, mediated by high energy states mainly of
oxygen character. Such a picture represents a smooth evolution to
doped case from the superexchange interaction in the half-filled
limit, and can be formally established through analyzing the
fluctuations beyond the mean field theory.\cite{Magint} Our analysis
shows that the overall amplitude of the magnetic interaction can be
determined unambiguously only in the limit of large or vanishing
$t_{pp}$. Because $t_{pp}$ in the cuprates is in the intermediate
region, we have appealed to phenomenology to pin down the amplitude of
the magnetic interactions. Our general conclusion will be that, in the
metallic region the magnetic interaction is of moderate strength. It
is non-zero, but far from causing a magnetic instability.

The ${\bf q}$-dependence of the magnetic interaction can be
specified on general grounds. We find that, for the antiferromagnetic
interactions the ${\bf q}$-dependence is to a good approximation given
by Eq. ({\ref{jq}). Such a ${\bf q}$-dependence reflects the smooth
evolution from the superexchange interaction in the half-filled case.
It is derived, as in the half-filled case, from the absence of the
on-site term due to the exclusion of double occupancy as a result of
the strong on-site Coulomb correlations. We emphasize that, these
arguments refer to the character of high energy states, and do not
depend on the details of the low energy quasiparticle dispersion.
Indeed Eq. (\ref{jq}) qualitatively describe the ${\bf
q}$-structure of the magnetic interaction in both YBaCuO and LaSrCuO
systems, despite their very different Fermi surface shapes.

The renormalized band structure determines the bare dynamical spin
susceptibility associated with the quasiparticles$-$the Lindhard
function.

\begin{eqnarray}
\chi_o ( {\bf q}, \omega ) ~\sim ~ \sum_{\bf k}
{ f({\bf k}) - f ({\bf k+q}) \over
\omega - ( E ( {\bf k} ) - E ( {\bf k+q} )) + i \eta}
\label{lindhard}
\end{eqnarray}
where $E ({\bf k})$ is the quasiparticle energy dispersion. In our
approach, the Lindhard function has already incorporated substantial
amount of interaction effects, through the Coulomb renormalization of
the quasiparticle dispersion. It incorporates low energy scales
induced by interaction effects, in addition to those associated with
specific bandstructure such as van Hove singularity and nesting.
The energy scale associated with the mass enhancement is called the
coherence energy ${\rm T_{coh}}$.

For a magnetic interaction which is not too close to induce a magnetic
instability, the dynamical spin susceptibility has the generalized RPA
form,\cite{note4}

\begin{eqnarray}
\chi ( {\bf q}, \omega ) ~\sim ~
{\chi_o ( {\bf q}, \omega ) \over 1 - J ( {\bf q}) \chi_o ( {\bf q},
\omega )}
\label{chirpa}
\end{eqnarray}
In this way, the magnetic interaction induces further softening of the
antiferromagnetic fluctuations. When the magnetic interaction
approaches to leading to a magnetic instability, processes beyond RPA
will become important. This will be further discussed in Section
\ref{sec:discu}.

\bigskip
\section{\bf Theory of Spin Dynamics in the Metallic Cuprates:\\
Neutron Scattering and NMR/NQR Relaxation}
\label{sec:theory}

We now apply this formalism to the analysis of the spin fluctuation
spectra. We aim at understanding the qualitative aspects of the
experimental data.

\subsection{Wave Vector Dependence of the Dynamical Susceptibility in
{\rm LaSrCuO}}

In order to understand the neutron cross section, we start with a
discussion of the Lindhard function in two dimensions. As is
emphasized in Ref. \cite{Lu}, the (dynamical) Kohn anomaly in two
dimensions leads to peaks in $\chi '' ({\bf q}, \omega )$ at the wave
vectors ${\bf q}=2{\bf k}_F$, in contrast to three dimensional case.
This feature was emphasized independently in Ref. \cite{Littlewood}.
It has been further studied in Ref. \cite{Tremblay} within the one
band Hubbard model for weak to intermediate Coulomb correlations.
In an isotropic system, the Kohn anomaly gives rise to a ring of
peaks. For a general Fermi surface, the shape of the peaked region in
$\chi '' ({\bf q}, \omega )$ reflects the geometry of the Fermi
surface. When applied to LaSrCuO with a Fermi surface given in Fig.
\ref{fermisurface}(a), nesting-like effects
further enhance some portion of this peaked region. In fact, it
induces four absolute maxima, at ${\bf Q^*}={\pi \over a} ( 1 \pm
\delta, 1)$, ${\pi \over a} ( 1, 1 \pm \delta)$.\cite{Schulz} This is
clearly seen in the three-dimensional ${\bf q}$-structure shown in
Fig. \ref{sq214}(a). The value of the incommensurability, $\delta$,
reflects the nesting wave vector which depends on the relative
distances between the nearly flat regions of the Fermi surface:
$\delta$ increases when the nearly flat portions of the Fermi surface
are closer to the ${\rm \Gamma}$ point.

The magnetic interaction has a ${\bf q}$-dependence of the form Eq.
(\ref{jq}) and is peaked at ${\bf Q}_{\rm AF}=({\pi \over a},{\pi
\over a})$. The effect of the magnetic interaction is to enhance
$S({\bf q},\omega)$ around the whole region near ${\bf Q}_{\rm AF}$.
For moderate values of ${J_o / J_c}$, (where $J_c$ is
the strength of the interaction which gives rise to a magnetic
instability), the four peak structure is enhanced without changing
shape. This is clearly seen in the three-dimensional ${\bf
q}$-structure at $J_o / J_c = 0.6$ shown in Fig. \ref{sq214}(b). When
projected along two of the four peaks, as is shown in Fig.
\ref{sq214}(c), this can be directly compared to the experimental
result shown in Fig. \ref{neutronexpt}(b) (when resolution broadening
is taken into consideration).

We now turn to the implications of the experimentally found large and
strongly doping (x) dependent incommensurability ($\delta$) shown in
Fig. \ref{neutronexpt}(c). We plot in Fig. \ref{incomm} $\delta$ vs. x
calculated within the present three-band large $U$ model. For
comparison, we also show the corresponding results for the small $U$
Hubbard model with nearest neighbor hopping $t$ only, and with next
nearest neighbor hopping $t'$ in addition to $t$. For the small $U$
one band Hubbard model with only a nearest neighbor hopping, the
incommensurability vanishes at half-filling and is strongly doping
dependent: the nesting wave vector is equal to ${\bf Q}_{\rm AF}$ at
half-filling, and deviates from ${\bf Q}_{\rm AF}$ as doping is increased.
However, the incommensurability is smaller than the experimental value
by more than a factor of two. When an additional next nearest neighbor
hopping is included, the Fermi surface is ``twisted''. With
appropriately chosen $t'$, the flat regions on the Fermi surface are
further apart, and the incommensurability is enhanced. However, the
next nearest neighbor hopping in this case destroys the special
perfect nesting feature of the half-filled limit, and doping induces
only very weak modifications to the incommensurability.\cite{note}
This is again inconsistent with experimental results.

In the three band large U case, the incommensurability is enhanced for
the same reason that the flat regions of the Fermi surface are further
apart compared to the one band nearest neighbor tight binding case.
However, this change of the Fermi surface shape occurs due to the
oxygen dispersion (the oxygen-oxygen hopping $t_{pp}$), which couples
to the copper states through the renormalized hybridization
$V_{pd}^*$. In the strong U limit, the rigid band picture breaks down:
the renormalized hybridization $V_{pd}^*$ decreases as the doping
concentration is decreased. As a result, the amount of the Fermi
surface ``twisting'' gets smaller, and the incommensurability is
decreased. In the asymptotic half-filling limit, the renormalized
hybridization $V_{pd}^*$ is reduced to zero, and the antiferromagnetic
wave vector approaches the commensurate limit ${\bf Q}_{\rm AF}$.
Therefore, our results for the incommensurability is consistent with
{\it both features of the experimental data for incommensurability,
i.e. the relative large values and the strong doping dependence}.

We emphasize that, in our calculation of $S({\bf q}, \omega )$ at
different dopings, the bare parameters of the Hamiltonian  are fixed.
The change in the renormalized band structure as the doping
concentration changes is determined self-consistency from the large
$U$ constraint. Therefore, we argue that the experimentally observed
{\it large and strongly doping dependent incommensurate structure} is
a manifestation of the strong Coulomb correlation effect in the spin
fluctuation spectra in ${\rm LaSrCuO}$.

\subsection{Wave Vector Dependence of the Dynamical Susceptibility in
{\rm YBaCuO}}

We now proceed to study the spin dynamics in the YBaCuO family. The
${\bf q}$-structure of the Lindhard susceptibility corresponding to
the appropriate band structure with a Fermi surface of Fig.
\ref{fermisurface}(b) is shown in Fig. \ref{sq123}(a). Because of the
Fermi surface rotation, the Kohn anomaly induced$-$and nesting
enhanced$-$peaks are far away from the antiferromagnetic wave vector
${\bf Q}_{\rm AF}= ({\pi\over a},{\pi\over a})$.
Therefore, the dynamical susceptibility is essentially featureless
near ${\bf Q}_{\rm AF}$.

Since the magnetic interaction is peaked around ${\bf Q}_{\rm AF}$,
it enhances the amplitude of $S({\bf q},\omega )$ or $Im \chi ({\bf
q},\omega )$ in the region surrounding ${\bf Q}_{\rm AF}$.
This results in an essentially commensurate peak shown in Fig.
\ref{sq123}(b). When projected along the diagonal direction, shown in
Fig. \ref{sq123}(c), our results can again be directly compared to
experimental results such as shown in Fig. \ref{neutronexpt}(a). Our
assumption that $J$ is not particularly close to leading to a magnetic
instability yields a peak relatively broad with a width essentially
independent of temperature, as is illustrated in Fig. \ref{sq123}(c).

An exchange interaction peaked at antiferromagnetic wave vectors is
therefore crucial to our analysis. Such a ${\bf q}$-dependence for the
magnetic interaction occurs naturally in the strong coupling limit,
as emphasized in last section. This is very different from the weak
coupling limit: the dominant exchange interaction is simply the
on-site (${\bf q}$-independent) term $U$ for small $U$.

We conclude that, while the broad commensurate peaks of $S({\bf
q},\omega)$ in YBaCuO are in striking contrast with the sharp
incommensurate structure in LaSrCuO, they are in fact {\it a
consistency check} within our theory: a) the absence of the
incommensurate peaks in YBaCuO reflects the fermiology of this system,
as does the presence of the incommensurate peaks in LaSrCuO; b) The
presence of the commensurate peak in YBaCuO reflects an exchange
interaction peaking at ${\bf Q}_{\rm AF}$. This reflects the strong
Coulomb correlations, as does the large and strongly doping dependent
incommensurability in LaSrCuO.

A number of other cuprate families have rotated Fermi surfaces. Our
analysis leads to the prediction that spin fluctuations in these
systems will also be commensurate.

\subsection{Temperature and Frequency Dependence of the Dynamical Spin
Susceptibility}

The role of low energy scales has been the theme of many studies in
the metallic cuprates.\cite{Review} Low energy scales can arise from
features in the bandstructure such as the van Hove singularity
($\omega_{VH}$ defined above) and/or the proximity to nesting. They
can also occur due to the proximity to various instabilities; we call
the soft energy scale due to the proximity to the Mott localization
the coherent energy ($T_{coh}$) and that due to the proximity to the
magnetic instability the spin fluctuation energy ($\omega_{sf}$).

The  frequency and temperature dependence of the dynamical
susceptibility in the present theory for both YBaCuO and LaSrCuO are
discussed extensively in Ref. \cite{Si}. The lowest energy scale is
manifested as  a weak peak in the frequency dependence, and is found
to correspond to the van Hove singularity energy $\omega_{VH}$ defined
above. This energy scale is larger for YBaCuO than that for LaSrCuO.
At a moderate strength, the magnetic interactions slightly lower the
peak energy, and considerably enhance the low energy spin fluctuation
spectra.

Alternatively, the low energy scales can be illustrated in a scaling
plot of the susceptibility in terms of $\omega/T$. When a low energy
scale, say $E^*$, is present, the frequency dependence of the spin
fluctuation spectra depend on $\omega$ in terms of $\omega / E^*$ for
frequencies and temperatures smaller than $E^*$. On the other
hand, at higher frequencies and temperatures, the spin fluctuation
spectra can scale with $\omega / T$. A signature for a low energy
scale is a deviation from $\omega / T$ scaling at low temperature and
frequencies. In Fig. \ref{scaling}, such scaling plots are shown for
susceptibilities calculated for both LaSrCuO and YBaCuO systems. A
deviation from scaling is clearly seen in YBaCuO, when the frequency
is smaller than $\omega_{VH}$ {\it and} the temperature smaller
than $T_{VH} \sim {1 \over 4} \omega_{VH}$, where $\omega_{VH} \sim 25
meV$. This prediction is consistent with subsequent experimental
results of Ref. \cite{Tranquada}(a) for YBaCuO shown in Fig.
\ref{neutrontw}. The scaling behavior persists to lower energies in
LaSrCuO, since $\omega_{VH}$ is smaller.\cite{Ruvalds} Systematic
experimental studies in LaSrCuO is at this stage under way.\cite{comm}

\subsection{NMR/NQR Relaxation Rate}

Thus far, we have given a systematic presentation of the wave vector,
frequency, and temperature dependence of $\chi''({\bf q},\omega)$, as
is reflected in neutron scattering data for both YBaCuO and LaSrCuO.
We now turn to the analysis of the NMR/NQR data.

Shown in Fig. \ref{nmr}(a) is the calculated temperature dependence of
the NMR relaxation rate at the copper site in YBaCuO case, which can
be directly compared with the experimental data above ${\rm T_c}$
given in Fig. \ref{1t1expt}}. The results for LaSrCuO are similar.
The high temperature saturation in our
calculation occurs due to a proximity to Mott localization {\it and}
the moderate magnetic interaction effects. The proximity to Mott
localization is reflected in the moderate mass enhancement, leading to
a coherent energy scale ${T_{coh}}$ above which the spin fluctuation
spectral weight starts to saturate. The antiferromagnetic interaction
leads to a further softening of the spin fluctuation spectra.
The van Hove energy scale discussed in the context of
neutron data plays a relatively minor role in ${1/T_1}$ due to the
integrability of the logarithmic divergence.

Since considerable interaction effects are already incorporated in
$\chi_o$, our assumption of a {\it moderate strength} of the magnetic
interaction $J$ is consistent with both the experimentally observed
enhancement and the deviation from linear in temperature
dependence in $({1\over T_1})_{Cu}$. This is in contrast with weak
coupling approaches in which
the interaction has to be fine-tuned to be extremely close to leading
to a magnetic instability in order to yield soft-enough temperature
scale in the temperature dependence of ${1 \over T_1}$ and strong
enough enhancement in its magnitude.

The situation for $({1\over T_1})_O$ is more complex. Because peaks in
the the dynamical spin susceptibility is strongly {\it incommensurate}
in LaSrCuO, and nearly-commensurate but {\it broad} in YBaCuO, a
perfect form-factor cancellation is not expected in either system.
Indeed, as can be seen in Fig. \ref{nmr}(b) for the YBaCuO case,
$({1\over T_1})_O$ shows considerable deviation from the linear in
temperature dependence. We note that, experimental study of ${\rm
({1 \over T_1})_O}$ is at present less thorough than that for ${\rm
({1 \over T_1})_{Cu}}$. Recent data\cite{newt1} indicate that, over
certain temperature range, ${\rm ({1 \over T_1})_O}$ in nearly optimal
doped ${\rm LaSrCuO}$ may not be inconsistent with what is expected
 from an incommensurate spin fluctuation spectra as revealed in the
neutron scattering. The reconciliation of neutron and ${\rm ({1 \over
T_1})_O}$ deduced spin fluctuation spectra is among the most important
issues to be further studied, both experimentally and theoretically.

\bigskip
\section{\bf Further Discussion}
\label{sec:discu}

The spin dynamics in the cuprates is an extensive subject. We have
given a detailed description of certain aspects. In this section, we
sketch on some related issues.

\subsection{Additional Aspects of the Spin Dynamics in the Normal State}

In this subsection, we briefly comment on the double layer magnetic
coupling in YBaCuO, the temperature dependence of the uniform spin
susceptibility, and the related ``pseudo-gap'' effects in the
dynamical spin spectra.

One important question in the physics of the high ${T_c}$ copper
oxides is the role of interlayer couplings in the low energy
electronic dynamics. In this regard, one feature which has been
established experimentally is the double-layer magnetic coupling in
YBaCuO.\cite{bilayer} It is possible to address this problem
microscopically in terms of Coulomb renormalizations.
As a first step towards understanding the phenomenological
implications, we have carried out an RPA analysis of this double-layer
magnetic correlation in Ref. \cite{Zha1}(b). Incorporating an
interplayer electronic coupling ($t_{\perp}$) extracted from the
splitting in the ARPES-induced Fermi surfaces\cite{ARPES} in ${\rm
YBa_2Cu_3O_{7-\delta}}$, we find that both the frequency dependence
and the c-direction wave vector $q_z$ dependence of $S({\bf
q},\omega)$ can be understood provided an interlayer magnetic coupling
($J_{\perp}$) is present. The extracted $J_{\perp}$ and $t_{\perp}$
are comparable in magnitude, suggesting that non-perturbative effects
can be important.

Some classes of cuprates exhibit a gap-like  features in the magnetic
dynamics above ${T_c}$. This is most extensively studied in the
deoxygenated YBaCuO in which the spin susceptibility $\chi$ as well as
$({1 \over {T_1T}})_O$ and $({1 \over {T_1T}})_Y$ (and over a much
narrower temperature range $({1 \over {T_1T}})_{Cu}$) all decrease
substantially as temperature is decreased,\cite{pseudogap} and a
gap-like feature occurs in the frequency dependence of the dynamical
spin susceptibility\cite{Rossat-Mignod}. Along the line of approach
taken here, higher order processes beyond RPA become important when
the magnetic interaction drives the system close to an
antiferromagnetic instability. When these additional processes are
taken into account, a new energy scale can be generated below which
$\chi$ decreases as temperature is lowered.\cite{thesis} In addition,
specific quasiparticle dispersion can lead to a ``pseudo-gap'' in
$S({\bf q},\omega )$ for more extended ${\bf q}$ values.\cite{lavagna}
Detailed analysis is needed to see whether
the ``pseudo-gap'' features indicate additional non-perturbative
effects. A discussion of these ``pseudo-gap'' features in terms of
RVB singlet formation can be found in Ref. \cite{RVB}. Recently, it
has been proposed that these ``pseudo-gap'' features may be associated
with charge degrees of freedom which occur for all ${\bf q}$, and are
related to the short coherence length in the superconducting
states.\cite{mohit} Alternative arguments have been made that, they
are associated with the interlayer singlet formation between adjacent
${\rm CuO_2}$ layers within a unit cell.\cite{millis_monien}

\subsection{Spin versus Charge Dynamics in the Normal State}

One central issue in the metallic cuprates is the relation between
spin and charge dynamics. This is highlighted by the proposal of spin
charge separation.\cite{Anderson} Experimentally, the low energy spin
fluctuation spectra exhibit various low energy scales. As we have
discussed in detail, they are well described by the coherent spectra
associated with the correlated quasiparticles. In addition, the spin
fluctuation spectra exhibit crossover behavior at intermediate energy
range (which we loosely define as, say, a decade of scale around
100meV).\cite{Imainew} These features are strikingly similar with spin
fluctuation spectra in the heavy fermion systems.\cite{Review,loc} In
charge responses, however, low energy scales appear only in
non-optimally-doped cuprates while intermediate energy crossover does
not seem to occur, as demonstrated in the ubiquitous linear
temperature dependence in resistivity and, perhaps more strikingly,
the quadratic temperature dependence in the Hall angle over a wide
range of temperatures\cite{Ong}. At higher energies, not much is known
about spin dynamics in the metallic cuprates,
while various anomalous features in the charge dynamics have been
found through optical, Raman, and photoemission spectroscopy studies.
A systematic study comparing spin and  charge dynamics at various
energy scales should reveal much on the physics of the copper oxides.

In the half-filled Mott insulator, strong Coulomb correlations push
the spectral weight for charge excitations to higher energies, leaving
spin excitations to play a dominant role at low energies. It is
therefore natural to address how the low energy spin fluctuations
affect the low energy charge dynamics in the doped case. Several
groups have addressed the contribution of the spin fluctuations to
resistivity.\cite{resis,Resis} In the context of neutron data, we
mentioned that the lowest energy scale in the spin fluctuation spectra
within our model came from the two-dimensional van Hove singularity,
$\omega_{VH}$. In Ref. \cite{Resis}, we found that $\omega_{VH}$ is
small over a wide range of doping concentration due to the strong
Coulomb correlations. This leads to a linear in T dependence in the
spin fluctuation induced scattering rate, {\it over a wide temperature
range} above $T_{VH}$. Furthermore, since $\omega_{VH}$ plays only a
minor role in the NMR relaxation rate due to the integrability of the
logarithmic divergence, this is consistent with differences in the low
energy scales observed in different experimental probes. Further work
is needed to address whether the spin fluctuations alone can account
for all the anomalous low energy charge responses, or additional
anomalous charge excitations\cite{Anderson,MFL} have to be invoked.

\subsection{Spin Dynamics in the Superconducting State}

Considerable attention has focussed recently on the spin dynamics
below ${\rm T_c}$. The central issue has been the extent to which one
can infer, from the spin fluctuation spectra, the nature of the
superconducting order parameter in the system: anisotropic versus
isotropic pairing and the existence or not of nodes at the Fermi surface.
Theoretical interpretations of the Cu and O Knight shifts and NMR
relaxation rates in ${\rm YBa_2Cu_3O_7}$ when combined with
anisotropy studies\cite{Barrett} have been argued to give strong support
for a $d_{x^2-y^2}$
pairing state.\cite{Bulut2,JP} In addition, recent neutron data on
both ${\rm La_{1.85}Sr_{0.15}CuO_4}$ and ${\rm YBa_2Cu_3O_{6.6}}$
reveal temperature T and frequency $\omega$ dependent features which
show considerable spin fluctuation spectral weight within the
superconducting gap.\cite{Mason,Thurston,Tranquada}

Taking into consideration the interplay between fermiology and the
Coulomb correlation effects, we have found that\cite{Zha2} the
anomalous temperature dependences at low frequencies observed in
neutron measurements\cite{Mason,Thurston,Tranquada} of $S ({\bf
q},\omega )$ are compatible with a  $d_{x^2-y^2}$ pairing state in
both LaSrCuO and YBaCuO. The calculated change in the ${\bf
q}$-structure of $S ({\bf q},\omega )$ as one goes from above to below
${T_c}$, due to a $d_{x^2-y^2}$ order parameter\cite{JP}, is expected
to occur most strikingly in YBaCuO. The lack of such a change in ${\bf
q}$-structure in recent experiments in both YBaCuO\cite{Tranquada} and
LaSrCuO\cite{aeppli} poses a serious question to the anisotropic
pairing scenario.

\bigskip
\section{\bf Conclusions}
\label{sec:conclu}

The spin fluctuation spectra in the normal state of the metallic
cuprates exhibit various anomalies. The contrasting wave vector
dependence of the dynamical spin structure factor for YBaCuO and
LaSrCuO, as well as their anomalous temperature and frequency
dependence, provide the basis for our argument that, both the
fermiology and strong Coulomb correlations play important roles in the
low energy spin dynamics.

We have shown that, the existence in LaSrCuO and the absence in YBaCuO
of incommensurate antiferromagnetic peaks are natural consequences of
the different shapes of the Fermi surfaces. At the same time, both the
large value and the strong doping dependence of the incommensurability
in LaSrCuO, and the presence of the commensurate peaks in YBaCuO, are
the manifestation of the strong Coulomb correlation effects (along
with fermiology effects). Within our analysis, the strong Coulomb
correlations lead to two effects. It renormalizes the quasiparticles,
and induces residual exchange interactions. Because of the
quasiparticle renormalization, we find that there can be enough spin
fluctuation spectral weight to explain the temperature and frequency
dependence in the neutron and NMR relaxation rate, with an
antiferromagnetic interaction which is far from being on the verge of
leading to a magnetic instability. Meanwhile, the various energy
scales found in the resulting spin fluctuation spectral function are
consistent with energy scales within the renormalized coherent
quasiparticle description of the spin dynamics.

We are thus led to conclude that, the low energy spin dynamics of the
metallic cuprates are appropriately described in terms of correlated
quasiparticles with a Luttinger Fermi surface, in contrast to
excitations in a frustrated spin system. These quasiparticles interact
with a moderate value of residual antiferromagnetic interaction. These
conclusions place strong constraints on the issue of the relevance of
the antiferromagnetic spin fluctuations to superconductivity.\cite{sc}
They also provide a prototype for the low energy spin dynamics in the
doped Mott insulators.

\acknowledgments

This brief review is based on works done in collaboration with Yuyao
Zha and K. Levin, and earlier collaboration with J.P. Lu and Ju Kim. I
would like to thank K. Levin for valuable comments on the manuscript,
and G. Aeppli, S.-W. Cheong, G. Kotliar, P. Littlewood, J. Tranquada,
R. Walstedt, and J.J. Yu for useful discussions. This work was
supported by the NSF under grant DMR 922-4000.

\figure{Spin lattice relaxation rates a) at the planar copper and
oxygen sites in ${\rm YBa_2Cu_3O_{7-\delta}}$ (with ${\rm
T_c=93K}$)\cite{Hammel} and b) at the planar copper site in ${\rm
La_{1.85}Sr_{0.15}CuO_4}$\cite{Imai}. \label{1t1expt}}

\figure{Neutron scattering cross section as a function of momentum
${\bf q}$ a) in ${\rm YBa_2Cu_3O_{6.6}}$ at T=10K for a diagonal scan
${\bf q}=h {2\pi \over a} (1,1)$ (The ${\bf q}$ dependence for $T>T_c$
is essentially the same. From Ref. \cite{Tranquada}(a)); and b) in
${\rm La_{1.86}Sr_{0.14}CuO_4}$ at T=35K for a scan ${\bf q}={\pi
\over a} (Q_x+{\delta \over 2},Q_x-{\delta \over 2})$, i.e. along two
of the four maxima at ${\bf Q^*}={\pi \over a} (1 \pm \delta, 1)$,
${\pi \over a} (1, 1 \pm \delta)$ where $\delta=0.245$ (from Ref.
\cite{Mason}(a)). Plotted in c) is the incommensurability versus the
doping concentration (from Ref. \cite{Mason}(b)).\label{neutronexpt}}

\figure{The ${\bf q}$-integrated dynamical spin susceptibility deduced
from neutron scattering cross section in ${\rm YBa_2Cu_3O_{6.6}}$ as a
function of $T/\omega$. The arrows show the temperature below which
$\chi''(\omega)$ deviates from $T/\omega$ scaling. Data of Ref.
\cite{Tranquada}(a).\label{neutrontw}}

\figure{Plasma frequency $\omega_p$ as a function of hole
concentration x a) calculated from the renormalized quasiparticle
band\cite{Si} and  b) derived from Drude-fitting the optical
spectra\cite{optical}. The ``fitted'' curve in b) represents $\omega_p
\sim \sqrt{x}$. \label{plasma}}

\figure{The Fermi surface of the renormalized quasiparticle band for
a) ${\rm La_{1.85}Sr_{0.15}Cu O_4}$ and b) ${\rm YBa_2Cu_3O_7}$ (from
Ref. \cite{Si}). The LDA-calculated Fermi surface is shown for c)
${\rm La_{1.8}Sr_{0.2}Cu O_4}$ (from Ref. \cite{LDA}(a)) and d) ${\rm
YBa_2Cu_3O_7}$ (shaded areas, Ref. \cite{LDA}(b)). Open circles in d)
correspond to the ARPES-measured Fermi surface for ${\rm
YBa_2Cu_3O_{6.9}}$ (from Ref. \cite{ARPES}(b)). \label{fermisurface}}

\figure{Calculated $S({\bf q}, \omega )$ versus ($ q_x , q_y$) for
${\rm La_{1.82}Sr_{0.18}CuO_4}$ a) from the Lindhard function
($J_o=0$) contribution; and b) for $J_o/J_c=0.6$; Also shown is c) the
projection of b) along two maxima, with ${\bf q}={\pi \over a}
(\kappa+{\delta \over 2},\kappa-{\delta \over 2})$ and $\delta=0.34$.
Here the temperature and frequency are 1 and 10 $meV$ respectively.
{}From Ref. \cite{Lu}(b).\label{sq214}}

\figure{Comparison of the incommensurability $\delta$ as a function of
doping concentration x in one band models (dashed line, $t_2=0$ and
dot-dashed line $t_2/t_1 = 0.25$) with the present theory (solid
circles). From Ref. \cite{Si}.\label{incomm}}

\figure{Calculated $S({\bf q}, \omega )$ versus ($ q_x , q_y$) for
${\rm YBa_2Cu_3O_{6.7}}$ a) from the Lindhard function ($J_o=0$)
contribution; b) for $J_o/J_c=0.7$. Here the temperature and frequency
are 1 and 10 $meV$ respectively. Also shown is c) the projection of b)
along the diagonal direction. From Ref. \cite{Zha1}(a).\label{sq123}}

\figure{Normalized $\chi''( {\bf q} , \omega )$ as a function of
$\omega /T$ for a) ${\rm La_{1.82}Sr_{0.18}CuO_4}$ and b) ${\rm
YBa_2Cu_3O_7}$. The solid lines represent the scaling curves. From
Ref. \cite{Zha1}(b). \label{scaling}}

\figure{Temperature dependence of NMR relaxation rate in ${\rm
YBa_2Cu_3O_{7-\delta}}$ (a) at the copper site and b) at the oxygen
site. From Ref. \cite{Si}. \label{nmr}}


\begin{references}
\bibitem[46]{note} Here the incommensurate peaks are identified as the
branch due to the nearly-nesting feature which is analytically
continued to that in the $t'=0$ case.
\end{references}
\end{document}